\documentclass{article}
\usepackage{spconf,amsmath,epsfig}
\usepackage{colortbl}
\usepackage{xcolor}
\usepackage{amsfonts}
\usepackage{bm} 
\usepackage{multirow}
\usepackage{amssymb}

\let\OLDthebibliography\thebibliography
\renewcommand\thebibliography[1]{
  \OLDthebibliography{#1}
  \setlength{\parskip}{0pt}
  \setlength{\itemsep}{0pt plus 0.3ex}
}

\begin{document}\sloppy

\def\x{{\mathbf x}}
\def\L{{\cal L}}

\title{Annotation-free Automatic Music Transcription
with Scalable Synthetic Data and Adversarial
Domain Confusion}
\name{Gakusei Sato$^*$ and Taketo Akama$^*$}
\address{Sony Computer Science Laboratories, Tokyo, Japan}
\maketitle
\def\thefootnote{*}\footnotetext{These two authors contributed equally to this work}

\begin{abstract}

Automatic Music Transcription (AMT) is a vital technology in the field of music information processing. Despite recent enhancements in performance due to machine learning techniques, current methods typically attain high accuracy in domains where abundant annotated data is available. 
Addressing domains with low or no resources continues to be an unresolved challenge. 
To tackle this issue, we propose a transcription model that does not require any MIDI-audio paired data through the utilization of scalable synthetic audio for pre-training and adversarial domain confusion using unannotated real audio.
In experiments, we evaluate methods under the real-world application scenario where training datasets do not include the MIDI annotation of audio in the target data domain. Our proposed method achieved competitive performance relative to established baseline methods, despite not utilizing any real datasets of paired MIDI-audio.
Additionally, ablation studies have provided insights into the scalability of this approach and the forthcoming challenges in the field of AMT research.
\end{abstract}
\section{Introduction}
\label{sec:intro}

Automatic Music Transcription (AMT) is a significant technology in the realm of music information processing. Generally, it seeks to convert music audio into a symbolic representation automatically, typically in the form of musical notes, pitch, timing, velocity, and other pertinent musical elements.

Constructing a practical AMT system necessitates a model adaptable to any instrument, a challenging task due to the complex overtone structures and dynamic articulations unique to each instrument and playing style.
The current most effective approach to solve this complex problem is supervised learning with large-scale models \cite{pianotransformer, mt3}. While various studies have been conducted and have contributed to significant performance improvements, this approach generally requires large annotated datasets.
However, as of now, no dataset exists with an adequate amount of data and diversity to tackle the complexity of such AMT tasks.
Therefore, most existing methods either specialize in a specific instrument type, like the piano, for which datasets are plentiful \cite{pianotransformer, hFT}, or they aim to facilitate multitasking by amalgamating existing datasets \cite{mt3, basicpitch, perceiver}.
While this multitasking approach has seen some success, data-rich domains like piano still consistently yield strong results \cite{mt3}. This performance may rely on the quality and quantity of currently available datasets and fails to achieve genuinely general performance across data domains not used in training.

One primary reason for the dataset shortage is the challenge of annotation. Supervised learning for AMT necessitates ground truth MIDI data corresponding to the audio, detailing the timing and pitch of the played notes.
The most straightforward methods to obtain this data are through expert human annotation or by capturing MIDI-audio data simultaneously with a MIDI instrument.
However, neither method is efficient, and many instrument domains still lack a substantial collection of annotated data for transcription \cite{annotation}. Additionally, the recent music industry has witnessed a diversification of tones in songs, owing to the rise of Desk Top Music (DTM) and other technologies. 
Merely amalgamating existing annotated data offers limited scope for enhancing generalization.

To solve this problem, we propose a transcription model with adversarial domain confusion of scalable synthetic and real domain data, achieving \textbf{annotation-free AMT} where the model does not require any MIDI-audio paired data for training. 
The domain confusion used here is a method of learning to make it difficult for a model to distinguish between different target domains.
By applying this method to synthetic domain audio and real domain audio, we fine-tune the transcription model pre-trained in the synthetic domain to the real domain.
In this process, synthetic domain data is generated by rendering MIDI data and one-shot audio data, and we only use unannotated audio data for real domain data. This enables a workflow that completely eliminates the need for MIDI-audio paired data.
In addition, we propose a scalable audio synthetic method using mixing of the one-shot audio. By using this method, we can train models with complex and diverse input patterns.

We conduct experiments in the \textbf{target-annotation-free setting} where training datasets do not include the MIDI annotation of audio in the target data domain.
The scores of our proposed method are competitive with those of established baseline methods, which utilize annotated data from other real domains, even though we do not use any MIDI-audio paired datasets for training.
We also perform ablation studies to examine the relationship between the timbre and MIDI variations of synthetic audio and their impact on transcription performance.

\section{Related work}
\subsection{Automatic Music Transcription (AMT)}
AMT has been extensively studied and researched.
It started with signal processing approaches like classical NMF \cite{NMF}, and in recent years, machine learning-based approaches have led to significant performance improvements \cite{overview}.
Particularly recently, large-scale networks utilizing transformer models have demonstrated high performance \cite{pianotransformer,mt3,hFT}.
Because of the aforementioned lack of a generic dataset, many studies have focused their approach on specific instrument types such as piano \cite{pianotransformer,hFT,pianosynth}, guitar \cite{gtsynth} and so on.
As approaches to multitasking, some studies have combined multiple datasets to train a single model \cite{mt3,perceiver}.

Several approaches have been developed to address less annotated (low-resource) domains.
Cheuk et al. used unannotated music recordings in semi-supervised learning for AMT \cite{reconvat}.
Wu et al. proposed a hybrid-supervised framework, the masked frame-level autoencoder (MFAE) to understand generic representations of low-resource instruments \cite{MFAE}.
Bittner et al. proposed an ``instrument-agnostic" method aimed at achieving general performance, independent of instrument types \cite{basicpitch}.
Cheuk et al. solved the transcription task using diffusion generative models, which boosts performance by pretraining on pianoroll data \cite{diff}.
Contrary to previous work, we propose a model that does not rely on real MIDI-audio paired datasets and demonstrates effective performance in the target-annotation-free setting.

\subsection{Synthetic audio}
In the field of music information processing, using synthesized audio in training is valuable for several reasons, particularly its ease of acquisition.
One method involves synthesizing MIDI datasets using one-shot audio to present a dataset specifically designed for instrument retrieval tasks \cite{nlakh}.
Slakh \cite{slakh} is an audio-MIDI paired dataset that renders music MIDI data with 187 different timbres using professional-grade sample-based virtual instruments, constituting the largest dataset suitable for transcription.

Beyond Slakh, several studies have also employed synthetic audio for AMT.
Kim et al. utilized synthetic audio generated by oscillators to augment the guitar transcription datasets \cite{gtsynth}.
Maman et al. utilized synthesized and unaligned piano audio data for the expectation maximization framework to enable training on in-the-wild recordings \cite{pianosynth}.
Simon et al. adopted a strategy of creating annotated polyphonic sounds by using monophonic sounds and their transcription results \cite{monopoly}.
In this paper, we generate synthetic audio by rendering various MIDI with one-shot audio of diverse tones.
This approach leverages the accessibility, scalability, and customizability of MIDI and one-shot audio.

\subsection{Deep domain adaptation and confusion}
In recent years, domain adaptation approaches that perform adversarial training to align the representation of source data with that of target data were proposed \cite{advda, advda2, advda3}.
Particularly, the technique of training models to learn domain-invariant representations is known as {\it domain confusion} \cite{domainconf}.
Tzeng et al. later proposed using a discriminator to achieve domain confusion \cite{advDC}.

In the context of music information processing, domain adaptation and confusion have been applied to various tasks. 
Bittner et al. studied microphone mismatch in automatic piano transcription using domain adaptation techniques \cite{DAmicrophone}. Lordelo et al. introduced adversarial domain adaptation to improve harmonic-percussive source separation in different domains \cite{DAseparation}. 
Narita et al. used domain confusion to extract pitch-invariant information, equating pitch differences with domain differences \cite{ganstrument}.
Mor et al. achieved music translation by domain confusion of sound features of different genres and different instruments \cite{univtrans}.
In this study, we demonstrate that employing domain confusion can make synthetic and real domains indistinguishable, thereby enhancing the transcription performance of real data.

\section{Model}
\subsection{Synthetic audio}
\begin{figure*}[t]
  \begin{minipage}[b]{1.0\linewidth}
    \centering
    \centerline{\epsfig{figure=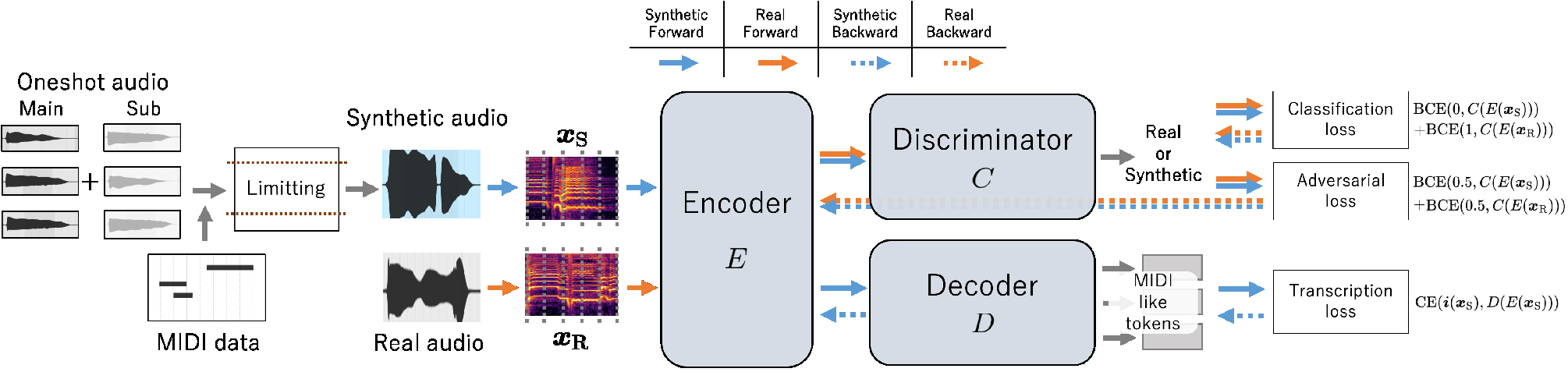, width=17.5cm}}
  \end{minipage}
  \caption{Our proposed method.}
  \label{fig:model}
\end{figure*}

The goal of this study is to build a transcription model achieving annotation-free AMT where the model does not require any MIDI-audio paired data for training. 
In this process, we generate the synthesized audio in a simple manner using MIDI data and one-shot audio data. 
First, to enhance timbre patterns, we mix audio samples as follows.
Two one-shot audio samples are randomly selected with different timbres from a one-shot dataset, where one is a sample of the main timbre ($T_{\mathrm{main}}$) and the other is that of the sub timbre ($T_{\mathrm{sub}}$).
For each pitch, samples of mixed timbres are calculated by $T_{\mathrm{main}} + \alpha T_{\mathrm{sub}}$.
The mixing rate $\alpha$ is randomized to increase timbre variation from a uniform distribution within the [0, 2] range, as determined by preliminary experiments. Following this process, we normalize the mixed one-shot audio to fall within a range of -1 to 1.
By doing this, $n(n-1)/2$ (+ difference of $\alpha$) different timbres can be created using $n$ different timbres. This has the effect of increasing the tonal diversity of the input data.

Next, this mixed one-shot audio and MIDI data are used to generate the synthetic audio.
For the instrument group of one-shot audio $T_{\mathrm{main}}$, the MIDI segment of the corresponding MIDI program number is chosen: 1-8 for keyboard, 9-16 for mallet, 17-24 for organ, 25-32 and 105-112 for guitar, 33-40 for bass, 41-56 for strings, 57-64 for brass, 65-72 for reed, 73-80 for flute, and 81-96 for both synth and vocal.
From the MIDI segment, we can obtain onset-time, offset-time, pitch, and velocity information for each note.
The one-shot audio is trimmed according to the note pitch and length (defined as offset-time minus onset-time) for each note.
If the length of a note exceeds the length of the one-shot audio, we simply use the full length of the audio.
The velocity ($v$) of each note determines the amplitude of each one-shot audio ($a$) by $a = \log(1+v/127)/\log2$.
Finally, audio samples for each note are overlaid and added together to produce the musical audio.

In general, detection of offset time is a difficult problem in transcription due to the release time.
The release time corresponds to the period following the offset time , which is not explicitly indicated by MIDI data.
Additionally, this varies across different instruments and is influenced by audio effects, like reverb.
In this study, the release time is defined randomly within a specific range based on the program number stored in the MIDI data: 0.1 to 1.0 seconds for keyboard, organ, mallet, brass, synth, and vocal, 0.8 to 1.0 seconds for strings, reed, and flute, 0.1 to 0.5 seconds for guitar, and 0.1 to 0.2 seconds for bass.

We also apply limiting to synthesized audio to simulate the dynamics that arise during the recording and production processes of real audio. With an 80\% probability, we set a randomly selected limiting threshold between 0.5 and 1.0 for normalized audio. Normalization is then reapplied after this step.

\subsection{Transcription architecture}
This section explains the transcription architecture as illustrated in Fig. \ref{fig:model}.
For input data, we use 2.56-second audio segments sampled in 16,000Hz, converted into mel spectrograms (window length is 2,048, hop length is 160, the number of mel bins is 384, and the time resolution is 10ms).  
Each time step of the mel spectrogram is fed into the fully-connected layer, the positional embedding is added, and then fed into the transformer as proposed in Gardner et al. \cite{mt3}.

For output data, we employ absolute time tokens based on MIDI-like representation \cite{miditok} and supplemented with an end-tie token as proposed in Gardner et al. \cite{mt3} to deal with notes that span separated audio segments. The tokens we use are pitch (128 values), time (256 values), note ON/OFF (2 values), begin of sequence/end of sequence (2 values), and end-tie (1 value). Note that velocity is not a prediction target in this study.
The vanilla transformer is employed as our transcription model \cite{transformer}.
Based on preliminary experiments, the transformer's internal parameters were set as follows: the dimension of the fully-connected layer is set to 1,024, the number of heads to 6, the number of encoder layers to 2, the number of decoder layers to 3, the embedding dimension to 384, and the learning rate to 1e-4.

After the pre-training, we perform domain confusion using synthetic audio data and real audio data as fine-tuning. 
First, both real and synthetic audio data are input into the transformer encoder. The encoder outputs from both synthetic and real data are transmitted into the discriminator, a three-layer fully-connected network, and the encoder output only from synthetic data is transmitted into the transformer decoder. 

Our goal is to bridge the gap between the synthetic and real domains in terms of timbral features, rather than the characteristics of note sequences.
We assume that the discriminator can focus on information other than the note sequence by inputting a short time range of the encoder output to the discriminator. To achieve this, 0.1-second intervals are randomly chosen from the encoder output and used as input to the discriminator.  
The learning rates for the transformer and the discriminator are set to 1e-5 and 1e-4, respectively.
As shown on the right side of Fig. \ref{fig:model}, the following objectives are alternately optimized:

\begin{align}
\underset{C}{\min}&\: \mathrm{BCE}(0, C(E(\bm{x}_{\mathrm{S}}))) + \mathrm{BCE}(1, C(E(\bm{x}_{\mathrm{R}}))),\label{q1}\\
\underset{E,D}{\min}&\: \mathrm{CE}(\bm{i}(\bm{x}_{\mathrm{S}}), D(E(\bm{x}_{\mathrm{S}})))\nonumber\\ 
+\lambda&(\mathrm{BCE}(0.5, C(E(\bm{x}_{\mathrm{S}}))) 
+\mathrm{BCE}(0.5, C(E(\bm{x}_{\mathrm{R}})))),\label{q2}
\end{align}

where $\mathrm{CE}$ denotes cross entropy loss and $\mathrm{BCE}$ denotes binary cross entropy loss. $\bm{x}_{\mathrm{S}}$ and $\bm{x}_{\mathrm{R}}$ are the input mel spectrograms of synthetic and real data, respectively. $E$, $D$, and $C$ are encoder, decoder, and discriminator, respectively. $\bm{i}(\bm{x}_{\mathrm{S}})$ is the one-hot vector indicating ground truth MIDI tokens of input $\bm{x}_{\mathrm{S}}$, and $0$ and $1$ are the ground truth labels of the input class ($0$ for synthetic and $1$ for real). 

Eq. \eqref{q1}, the classification loss, updates the discriminator's parameters to improve the discrimination performance between synthetic and real data. 
The first term in Eq. \eqref{q2}, the transcription loss, updates the  encoder and decoder's parameters to ensure that the decoder output approximates the correct MIDI token data, thereby improving transcription performance.
The second term in Eq. \eqref{q2}, the adversarial loss, updates the encoder's parameters to drive the discriminator output toward 0.5, making it more difficult to distinguish between synthesized and real data.
The parameter $\lambda$ In Eq. \eqref{q2} balances the effects of transcription and domain confusion, and was set to 0.01 through preliminary experiments.
\section{Experiment}

\begin{table*}[ht]
\centering \begin{tabular}{|c|c|c|c|c|c||c|c|c|c|c|c|} \hline
 &\multicolumn{3}{c|}{Real Data}&\multicolumn{2}{c||}{Our Synth Data}& Guitarset & Maestro & Molina & Phenicx & Slakh\\
&a & b & c &MIDI&Timbre& Fn\, Ac\, F & Fn\, Ac\, F & Fn\, Ac\, F & Fn\, Ac\, F & Fn\, Ac\, F \\
\hline
\hline
Bittner et al. \cite{basicpitch}&\checkmark&\checkmark&\checkmark&---&---&.79 .70 .56 &.71 .38 .11 & \cellcolor[rgb]{0.8, 1.0, 0.8}\textbf{.55} .60 \textbf{.38} & \cellcolor[rgb]{0.8, 1.0, 0.8}.51 .50 \textbf{.36} &.43 .40 .23\\
\hline
Wu et al. \cite{miamt}&\checkmark&\checkmark&\checkmark&---&---&\cellcolor[rgb]{0.8, 1.0, 0.8}.59 .43 .27 &.30 .39 .07 &\cellcolor[rgb]{0.8, 1.0, 0.8}.31 .48 .11 &\cellcolor[rgb]{0.8, 1.0, 0.8}.12 .13 .05 &\cellcolor[rgb]{0.8, 1.0, 0.8}.23 .40 .07\\
\hline
Gardner et al. \cite{mt3}&\checkmark&\checkmark&\checkmark&---&---& \cellcolor[rgb]{0.8, 1.0, 0.8}\textbf{.78}\, ---\, --- &\cellcolor[rgb]{0.8, 1.0, 0.8}.28\, ---\, --- & --- & --- &\cellcolor[rgb]{0.8, 1.0, 0.8}.14* ---\, ---\\
\hline
Simon et al. \cite{monopoly}&\checkmark&\checkmark&\checkmark&---&---&\cellcolor[rgb]{0.8, 1.0, 0.8}.71\, ---\, --- & \cellcolor[rgb]{0.8, 1.0, 0.8}\textbf{.83}\, ---\, --- & --- & --- &\cellcolor[rgb]{0.8, 1.0, 0.8}.45* ---\, ---\\
\hline
Real-Mix &\checkmark&\checkmark&&---&---&.80 .76 .58 &.90 .57 .54 &.78 .78 .58 &.66 .66 .40 &.79 .79 .60\\
\hline
Real-Omit &\checkmark&\checkmark&&---&---& \cellcolor[rgb]{0.8, 1.0, 0.8}\textbf{.78} .64 \textbf{.43} & \cellcolor[rgb]{0.8, 1.0, 0.8}.51 .47 \underline{.22} & \cellcolor[rgb]{0.8, 1.0, 0.8}.26 .66 .13 & \cellcolor[rgb]{0.8, 1.0, 0.8}.61 .51 .17 & \cellcolor[rgb]{0.8, 1.0, 0.8}.35 .42 .13\\
\hline
\hline
Synthetic-DC &\checkmark&&&126K&313K+& \cellcolor[rgb]{0.8, 1.0, 0.8}\underline{.77} \textbf{.67} \underline{.40} & \cellcolor[rgb]{0.8, 1.0, 0.8}\underline{.57} \textbf{.50} \textbf{.23} & \cellcolor[rgb]{0.8, 1.0, 0.8}.50 .66 .24 & \cellcolor[rgb]{0.8, 1.0, 0.8}\textbf{.74} .61 \underline{.32} &\cellcolor[rgb]{0.8, 1.0, 0.8}\textbf{.57} \underline{.57} \textbf{.31} \\
\hline
Synthetic-DA &\checkmark&&&126K&313K+& \cellcolor[rgb]{0.8, 1.0, 0.8
}.64 .60 .27 & \cellcolor[rgb]{0.8, 1.0, 0.8}.50 \underline{.49} .20 & \cellcolor[rgb]{0.8, 1.0, 0.8}.47 .67 .23 & \cellcolor[rgb]{0.8, 1.0, 0.8}.64 \underline{.63} .28 & \cellcolor[rgb]{0.8, 1.0, 0.8}.54 .54 .28 \\
\hline
Synthetic-L &&&&126K&313K+& \cellcolor[rgb]{0.8, 1.0, 0.8}.73 \underline{.66} .38 &\cellcolor[rgb]{0.8, 1.0, 0.8}.54 \underline{.49} \underline{.22} &\cellcolor[rgb]{0.8, 1.0, 0.8}\underline{.52} .67 \underline{.25} &\cellcolor[rgb]{0.8, 1.0, 0.8}\underline{.72} \textbf{.66} \textbf{.36} &\cellcolor[rgb]{0.8, 1.0, 0.8}\underline{.56} .55 \underline{.30}\\
\hline
Synthetic-M &&&&126K&792&\cellcolor[rgb]{0.8, 1.0, 0.8}.71 .65 .35 &\cellcolor[rgb]{0.8, 1.0, 0.8}.55 .47 .21 &\cellcolor[rgb]{0.8, 1.0, 0.8}.43 \underline{.68} .20 &\cellcolor[rgb]{0.8, 1.0, 0.8}.70 .60 .25 &\cellcolor[rgb]{0.8, 1.0, 0.8}\textbf{.57} \underline{.57} \textbf{.31}\\
\hline
Synthetic-S &&&&13K&792&\cellcolor[rgb]{0.8, 1.0, 0.8}.65 .63 .29 &\cellcolor[rgb]{0.8, 1.0, 0.8}.40 .41 .11 &\cellcolor[rgb]{0.8, 1.0, 0.8}.45 \textbf{.69} .20 &\cellcolor[rgb]{0.8, 1.0, 0.8}.68 .61 .30 & \cellcolor[rgb]{0.8, 1.0, 0.8}.56 \textbf{.61} .29\\
\hline
\end{tabular}
\caption{Results of the experiment. 
Scores highlighted in green pertain to the target-annotation-free setting, where the MIDI annotation of the target data domain for evaluation is not used for training to test the generalization capability. Among the scores under the target-annotation-free setting, we mark the best result with \textbf{boldface} and the
second-best result by \underline{underlining}. Our proposed method is \textit{Synthetic-DC}. \textit{Synthetic-DA}, \textit{L}, \textit{M}, and \textit{S} constitute an ablation study of the components integral to our proposed method, examining their individual and collective contributions to the model's performance.
`Real Data' `a', `b', and `c' each indicate whether the models are trained using real audio, real audio with MIDI annotation, and the extra real audio with MIDI annotation, respectively.
The asterisk (*) indicates the evaluation results for multitrack audio.
Our proposed method performs well under the target-annotation-free setting. The performance of synthetic methods tends to improve as the number of MIDI and audio variations used in the model increases.}
\label{tab:result}
\end{table*}

\subsection{Evaluation metrics}
We used mir-eval \cite{mireval} as the evaluation code, and three evaluation metrics were used. The first metric is the note-level F measure (\textit{F}), where a note is deemed correct if its pitch is correct, the onset error is within 50 ms, and the offset error is within 50 ms or 20 percent of the note length. The second metric is note-level F measure no offset (\textit{Fn}), evaluated similarly to F, but ignores offsets. The third metric is frame level accuracy (\textit{Ac}). 
Given the inherent challenges in offset estimation, this study primarily focuses on \textit{Fn}, following \cite{basicpitch}.

\subsection{Dataset}
The NSynth dataset \cite{nsynth} and Lakh MIDI dataset \cite{lakh} is used to synthesize the audio. The NSynth dataset contains large amount of 4 seconds one-shot samples with 11 instrument types, consisting of 1,006 different timbres, all collected from commercial sample libraries.
The Lakh MIDI dataset is a large MIDI dataset of 176,581 tracks collected from the Internet. We use the largest version, LMD-full, while omitting drums and percussion tracks

For evaluation purposes, we use Guitarset (guitar) \cite{guitarset}, Maestro (piano) \cite{maestro}, Molina (vocal) \cite{molina}, Phenicx (orchestra) \cite{phenicx}, and Slakh (mixture) \cite{slakh}.
In the case of Phenicx, we use instrumental grouped stems and midi. Note that Lakh MIDI dataset contains MIDI data for Slakh, but still no MIDI-audio pair in Slakh is used for our proposed synthetic method.

\subsection{Baselines}
Bittner et al. \cite{basicpitch}, Wu et al. \cite{miamt}, Gardner et al. \cite{mt3}, and Simon et al. \cite{monopoly} are selected as the state-of-the-art baselines, which perform the target-annotation-free setting where training datasets do not include the MIDI annotation of audio in the target data domain. Bittner et al. \cite{basicpitch} employed the same datasets for evaluation as ours, and the result for Phenicx is under the target-annotation-free setting. 
While they also did not use the Molina (vocal) dataset for training, they instead utilized the iKala (vocal) \cite{ikala} dataset.
Wu et al. \cite{miamt} trained the model with Maestro and MusicNet \cite{musicnet} dataset, so the results are target-annotation-free setting except for Maestro.
Gardner et al. \cite{mt3} and Simon et al. \cite{monopoly} demonstrated results in the target-annotation-free setting for Guitarset, Maestro, and Slakh. We only show \textit{Fn} scores for them because they did not use \textit{F} and \textit{Ac} for evaluation.
To demonstrate the effectiveness of our method with synthetic audio, we present the results of combining annotated real-domain datasets using the proposed transformer model. Here we do not apply domain confusion.
\textit{Real-Mix} is trained by combining five real audio datasets. \textit{Real-Omit} is trained by combining four real audio datasets without one real audio dataset of the target domain to create the target-annotation-free setting. 
For dataset combination, we adopt the balanced sampling scheme described in Gardner et al. \cite{mt3}.  If a dataset $i$ has $n_i$ examples, we sample examples from that dataset with probability $(n_i / \sum_j n_j)^{0.3}$.
Although these baselines and \textit{Real-Omit} do not use the target domain data for training, they utilize annotated real domain data for training. They may be operating under more favorable conditions compared to our proposed method that does not use MIDI-audio paired dataset.

\subsection{Proposed method}
Our proposed method is denoted as \textit{Synthetic-DC}. 
The model is trained using synthetic data and unannotated real audio data of the target domain. 
In an ablation study, domain adaptation is also conducted using the same transcription architecture, denoted as \textit{Synthetic-DA}.
While we train the model to make the discriminator output closer to 0.5 in domain confusion as in Eq. \eqref{q2}, we train the model to make the discriminator output closer to 1.0 in domain adaptation. 
Additionally, several ablation studies are conducted to evaluate the scalability of the synthetic audio.
\textit{Synthetic-L} uses Lakh MIDI dataset and mixed NSynth dataset, same as \textit{Synthetic-DC} except for domain confusion. \textit{Synthetic-M} uses Lakh MIDI dataset and non-mixed NSynth dataset. \textit{Synthetic-S} uses Slakh MIDI dataset and non-mixed NSynth dataset. \textit{Synthetic-L}, \textit{M}, and \textit{S} employ neither domain confusion nor domain adaptation. 

\subsection{Results}

The results of our proposed method and the baselines are shown in Table \ref{tab:result}. 
Scores highlighted in green pertain to the target-annotation-free setting, where the annotation in the target domain for evaluation is not used for training. Among the scores under the target-annotation-free setting, we mark the best result with \textbf{boldface} and the
second-best result by \underline{underlining}.
In `Real Data’ categories, `a’, `b’, and `c’ each indicate whether the models are trained using real audio, real audio with MIDI annotation, and the extra real audio with MIDI annotation, respectively.

Despite the absence of annotated real domain data, the scores for our proposed \textit{Synthetic-DC} are competitive with those of established baseline methods which utilize annotated real data. Within a group of strong baseline methods and those identified through ablation studies, our method frequently yields the best or second-best results in the target-annotation-free setting, thereby demonstrating consistently stable performance.
Comparing synthetic methods, \textit{Synthetic-DC} scores are higher for many datasets, indicating the effectiveness of domain confusion for transcription tasks. 
On the other hand, domain confusion tends to be less effective in bridging larger domain gaps, such as those found in the vocal dataset (Molina). This reduced efficacy is believed to stem from the unique structural characteristics of pronunciation.
In addition, comparing \textit{Real-Mix} and \textit{Real-Omit}, \textit{Real-Omit} scores tend to decrease significantly. This suggests that training by mixing multiple datasets, which has conventionally been used as an approach to multitasking, actually has weak generalizability to domains other than the ones used for training. In contrast, the proposed method consistently produces robust and stable results without using annotations of the target domain, showing a glimpse of generalizability.

For the \textit{Synthetic-L, M, and S}, performance tends to improve as MIDI and timbre variations increase.
While the result for the Maestro dataset shows significant performance improvement with increased MIDI variations, the result for the Molina dataset shows significant performance improvement with increased timbre variations. 
This implies that various instrument domains present distinct transcription challenges, and our synthetic approach is apt for adjusting the MIDI and timbre variations as required.

\section{Conclusion}
We proposed a transcription model that utilizes scalable synthetic audio and adversarial domain confusion to achieve annotation-free AMT which does not require any MIDI-audio paired data.
Through experiments on several datasets, we have shown that our proposed method performs well in the target-annotation-free setting. We also found that adversarial domain confusion using timbre information is useful for transcription.
The ablation studies indicate that distinct instrument domains present unique challenges, a critical consideration for future research targeting a universal model.

In this study, we utilized a public MIDI dataset and a public one-shot dataset for training.
However, such datasets are also published elsewhere, available for free or purchase, and can even be created through rendering with specialized software because they are often used in music production.
Additionally, our domain confusion fine-tuning method can be applied using only the target domain audio data without MIDI annotation.
We believe that by selecting and amalgamating a collection of such readily accessible datasets, we can construct models customized to particular objectives, thereby circumventing the need for manually annotating audio with MIDI for the specific domains we aim to transcribe.

\bibliographystyle{IEEEbib}
\bibliography{main}

\end{document}